\documentclass[a4paper]{jpconf}

\usepackage{graphicx}

\begin{document}
\title{Flexible Session Management in a Distributed Environment}

\author{Zach Miller$^1$, Dan Bradley$^1$, Todd Tannenbaum$^1$, Igor Sfiligoi$^2$}

\address{$^1$ University of Wisconsin, Madison, WI, USA}
\address{$^2$ Fermi National Acceleartor Laboratory, Batavia, IL, USA}

\ead{zmiller@cs.wisc.edu}

\begin{abstract}

Many secure communication libraries used by distributed systems, such as SSL,
TLS, and Kerberos, fail to make a clear distinction between the authentication,
session, and communication layers. In this paper we introduce CEDAR, the secure
communication library used by the Condor High Throughput Computing software,
and present the advantages to a distributed computing system resulting from
CEDAR's separation of these layers.  Regardless of the authentication method
used, CEDAR establishes a secure session key, which has the flexibility to be
used for multiple capabilities.  We demonstrate how a layered approach to
security sessions can avoid round-trips and latency inherent in network
authentication.  The creation of a distinct session management layer allows for
optimizations to improve scalability by way of delegating sessions to other
components in the system.  This session delegation creates a chain of trust
that reduces the overhead of establishing secure connections and enables
centralized enforcement of system-wide security policies.  Additionally, secure
channels based upon UDP datagrams are often overlooked by existing libraries;
we show how CEDAR's structure accommodates this as well.  As an example of the
utility of this work, we show how the use of delegated security sessions and
other techniques inherent in CEDAR's architecture enables US CMS to meet their
scalability requirements in deploying Condor over large-scale, wide-area grid
systems.

\end{abstract}

\section{Introduction}
CEDAR is the communications and security library used in the Condor
\cite{condor-hunter, condorgrid}
High-Throughput Computing software.  CEDAR is a multi-platform C++ 
TCP and UDP network socket representation that
offers several important features essential in a distributed
environment.  The cross-platform implementation allows a wide variety of
architectures to easily communicate over a network while maintaining a very
strong focus on security.  
CEDAR offers several improvements to overcome
shortcomings with traditional UDP datagrams, and makes timeout management and
error propagation first-class citizens which is essential in distributed
computing.  In this section we briefly introduce the salient CEDAR feature
set.

\subsection{Cross-platform Capabilities}
Basic communication principals are provided for sending integers, floating
point values, strings, ClassAds \cite{mms}, and even files, all in an
architecture-independent representation which allows for differing sizes of
core data types like integers and floating point values.  Integers on different
platforms may be different sizes, such as 32 vs. 64-bit, and so are padded and
sign-extended.  Also, CEDAR is aware of different byte ordering on different
platforms and automatically translates from one to another when necessary.
Furthermore,
floating point values are translated between single and double precision.  The
ability to send and receive ClassAds as first-class entities is vital to
CEDAR's use in Condor as ClassAds are the lingua franca of the Condor HTC
system \cite{gridbook-htc}, and allows for great flexibility in wire protocols.

\subsection{UDP Capabilities}
Traditional UDP datagrams have many shortcomings that are addressed by CEDAR to
increase UDP's usefulness in a typical distributed environment.  CEDAR does not
rely on the kernel to do packet fragmentation and reassembly, which has several
advantages.  Doing fragmentation and reassembly in user-space avoids using
precious system kernel memory which is often a fixed size and relatively
limited resource, thus allowing a higher volume of UDP traffic without dropping
packets.  Furthermore, it allows for datagrams greater than 64 kilobytes
(KB) in size, a
UDP
limitation that is easily exceeded by today's workloads.  For example, a single
ClassAd representing a job's environment could easily exceed 64 KB in size.
Another advantage to CEDAR's packet fragmentation is that CEDAR can optimize
the size of the fragments and introduce delays when sending them, which  we
have observed on many
operating systems results in fewer lost packets.  Finally, CEDAR adds 
security capabilities to UDP traffic, including authentication, encryption, and
integrity checks, which for UDP are often overlooked in other security layer
implementations such as SSL \cite{ssl}.

\subsection{Timeout Management}
In a distributed system where many components are communicating with one
another, it is essential that the amount of time spent blocking on I/O
operations is minimized, particularly if some of the connections are
established over Wide Area Networks (WANs) as is typical in a grid environment.
For this reason, CEDAR allows for carefully managing the timeouts for all I/O
operations so that one process is never stuck indefinitely waiting for another.
CEDAR does use non-blocking connects to avoid having processes sit idle, but to
avoid having one component of the system wreak havoc in other parts, timeouts
are used to break out of blocking I/O reads and writes in the event that one
component has failed or is lagging behind for any reason.

\subsection{Error Stack}
Because a single failure at a low level can often cause a cascade of error
conditions, CEDAR keeps an \emph{error stack} instead of a single integer
\emph{errno} which is the conventional approach, thereby
providing a richer method of propagating and reporting failures.  This
allows very detailed information on the root cause of the problem to be kept
while each layer chooses to either handle the error and continue, or to add its
own information to the error stack and propagate the stack to the next layer.
Ultimately, this leads to a more detailed and hopefully more informative message that
makes understanding both the high-level problem and low-level problem clear.

As an example, suppose a user was trying to query a Condor daemon and the
security settings are setup to require GSI \cite{globus-security} authentication.  However, the user
does not have a valid proxy.  Instead of a single ``111 -- connection refused''
error, the user would get something more like:
\begin{quote}
	111  -- connection refused\\
	1003 -- authentication failed\\
	5003 -- GSI: unable to acquire credential\\
	851968 -- GLOBUS: no valid proxy file
\end{quote}

Also, if an error is recoverable, any layer may pop the error off the stack,
take corrective action, and continue.  

\subsection{Basic Security Features}
Many basic security features essential for a distributed system are provided by CEDAR, including
authentication, encryption, and integrity checks.  There is also a layer for
the negotiation of which security features are available and allowed for a
given connection.  Finally, there is a layer which maps users from their native
credential to a canonical Condor identity and allows for authorization based on
this identity.  Currently in CEDAR the available authentication methods include
authentication using the local file system, SSPI \cite{sspi} (for Windows),
password, Kerberos \cite{kerberos} (via Active Directory on Windows), GSI
\cite{globus-security}, and OpenSSL \cite{openssl}.  This
variety of mechanisms allows strong authentication even between wildly different
operating systems such as Linux and Windows.  Integrity checks can by done with
MD5 \cite{md5} while encryption can be done with either 3DES \cite{3des} or
Blowfish \cite{blowfish}.  Current development efforts are adding AES and SHA-1
support.

When negotiating security features for a connection, it is the server (the side
which receives the incoming connection) who has final say over which features
are allowed and what the order of preference is.  It is also the server's
responsibility to perform the mapping from a security credential to
canonical identity (user name), and
inform the client who it ultimately was authenticated as.

\section{Importance of Session Management}

Capable network session management is critical to a highly scalable
distributed system that needs secure connections.  To understand why,
consider OpenSSL and GSI.  Both are authentication systems that make use of X.509
certificates, based on public-key infrastructure (PKI) technology.  While
they offer a cryptographically
strong authentication, there are two significant drawbacks to using these
methods exclusively: CPU overhead and network latency.  The first has
minimal impact with today's processors, but even a modest amount of CPU usage can become a bottleneck
as the scale of a distributed system increases.  For the
\emph{condor\_collector} service,
which handles incoming connections from all compute nodes in a cluster,
performing GSI authentication for many thousands of incoming connections can
overwhelm a server.  The latter concern of latency is a bigger issue, as quite
often the network connections in a grid environment are established across a
WAN.  Authentication using GSI or OpenSSL requires several round
trips across the network, which when combined with typical latencies over a WAN
can result in delays on the order of 1/10 to 1/4 of a second.

Kerberos is another authentication protocol supported by Condor, but it also
has potential bottlenecks.  Kerberos relies on network connectivity to the
centralized KDC service \cite{kerberos},
which again will introduce latency and even worse, can become overloaded due to
too many authentications happening.  Therefore, we need to minimize the number
of actual authentications performed in order to enable a reasonable level of
scalability.

This can be done by using a session management layer, which allows for a
semi-permanent information exchange that is explicitly set up and torn down.
The authentication is performed only once during the setup part, during which
secret keys are also securely distributed to both the server and the client.
These keys are then used to resume the session in the future, thus allowing the
high cost of authentication to be paid only once for the duration of the
conversation.

\subsection{CEDAR Sessions}

CEDAR implements its own session management layer, because of the delegation
opportunities enabled with this approach (see below), and because we found session
handling in other security libraries to be lacking.  For instance, OpenSSL
also provides session management, but cannot support sessions established
via authentication mechanisms other than SSL.  In addition, OpenSSL alone
also does not address the authentication of UDP packets and the ability to
use sessions over UDP.  Finally, OpenSSL sessions do not store as much state
as Condor would like, such as the mapped canonical name and the
authorization policy used to establish the connection.

The CEDAR session management layer is highly stateful and records several
important pieces of information about the session, including who the connection
claims to be, where it originated from, how it was authenticated, which
canonical user that credential maps to, how that user was authorized, and which
types of encryption and integrity checks will be used for the connection.
Using this extra data, many optimizations can be made, allowing the same
session to be used for other purposes as long as the new purpose is also
authorized by the policy used to create the original session.  This further
reduces the number of authentications and improves scalability.

After two entities make initial contact and authenticate each other with
whatever method was negotiated, a session is created and an internal 192-bit
session key is generated and distributed securely to each entity.  It is the
knowledge of this key that allows sessions to be resumed quickly and securely
without re-authenticating each time.  The keys have expiration dates after which
the session is torn down and a new one must be created.  The default lifetime
for a session varies depending on the purpose of the connection, but is also
configurable, and adjusting the session lifetime can impact the memory
footprint of a running daemon by limiting the number of sessions cached in
memory at any given point.  A shorter lifetime will cause fewer sessions to be
cached, which reduces the memory footprint but also increases the number of
authentications performed to periodically reestablish new sessions.

\subsection{Resuming a Session}
Condor's session management is optimized for the common case of resuming an
already established session.  When resuming a session, Condor sends the session
ID (and not the key itself of course) as part of the initial contact.  Each
entity looks up the session key using the session ID.  The sending party uses
the secret key to encrypt either the entire stream or at the very least the MD5
sum.  The ability of the receiving side to decrypt the stream or MD5 sum proves
they share the same secret key and therefore authenticates the resumed
connection.  Condor does not wait for confirmation that the session ID is
valid, as this would require a network round trip.  By optimizing for the
common case, there are no additional round trips in resuming a session, and
eliminating the round trip means there is no additional latency.  This makes
resuming a session very efficient, which is extremely important in constructing
a scalable distributed system where the network connections are potentially
made over a WAN with relatively high latency.

\subsection{Invalid Sessions}
Because Condor is optimized for the common case, in the event of an invalid
session being resumed the server must notify the client out-of-band that the
session is no longer valid.  This instructs the client to destroy the invalid
session and create a new one.  This could happen for instance if the server was
restarted for some reason (thus clearing the server's soft-state session cache) and the
client is not.  This message to invalidate a session is sent using TCP by
default since UDP packets are unreliable, but delivery of this information is
critical.

\section{Delegated Sessions} 

An important characteristic of CEDAR's session layer is the clean seperation
between network authentication and the establishment of a security session.  In
CEDAR, the authentication process results in the creation of a security
session.  This session contains all state required to perform secure
communications.  Furthermore, this session state is serializable and
transferable.  By transfering the session state from one service to another
that have established a trust relationship, a system using CEDAR is able to
establish a secure communication channel between two entities that had not
previously communicated.

For example, when the Condor system is matching jobs from a submit node to an
execute node, the centralized Condor directory service must already communicate
with both the nodes to perform matchmaking.  By delegating
the security session through these existing secure connections, Condor can establish a
secure channel between the submit and execute nodes even though they have not
previously authenticated to each other directly.  This then allows them to
communicate by resuming an existing session, the optimized common case, and
avoid performing the expensive authentication.  This is especially helpful
because the queueing service on the submit node is typically where scalability
is limited, and the limiting factor was often high-latency network
authentications and somewhat heavy CPU utilization due to PKI.  Delegating
sessions enables this cost to be offloaded to a third-party machine.

Another advantage to delegating trust relationships in this manner is it
allows two daemons to establish a secure channel even if they do not share a
common authentication mechanism between them.  For instance, GSI has not yet
been implemented on Windows despite GSI being widely deployed on world-wide
physics grids.  By having the central directory service delegate sessions to
the submit and execute nodes, systems with no common authentication methods can
still be part of the same grid.

\subsection{Advantages for UDP}
Most common communication libraries overlook the implementation of security
features for UDP datagrams.  Enabling security features via UDP can be a
challenge because it is connectionless and also has no delivery guarantees.
However, because of CEDAR's flexible session management, a session set up on
one connection can be reused for another connection, which allows CEDAR to
provide the ability to send and receive secure UDP datagrams.  A session must
first be established using TCP, since that is the only way many authentication
mechanisms can operate.  If a session already exists between two communicating
entities, there is then no additional overhead to using secure UDP.  If one
does not exist, CEDAR will automatically establish a session using TCP,
incurring only the minimal overhead of a single authentication and secure key
exchange.  After establishing a session, CEDAR can then use the session to send
secure UDP messages using the shared secret key in exactly the same way as a
TCP session is resumed and authenticated.  This technique adds no additional
round trips on the network and thus sending a UDP datagram remains a one-way
operation.

\subsection{Experimental Results} 
The US CMS collaboration is using Condor glideins\cite{glideinwms} to harvest
Grid resources; the Condor collector and schedd are kept in a central location,
while the Condor startds are sent to the Grid resources to create a
virtual-private Condor pool.  GSI authentication is used for all inter-daemon
communication.

The initial deployments used Grid resources located in the same region as the
central services and the results were very good; Condor could easily handle
10000 glideins, with a 1 Hz global glidein turnaround rate (equal to the global
glidein startup rate).  However, when testing the scalability on Grid resources
spread world-wide, Condor failed to scale beyond a few hundred glideins. The
network latencies were determined to be the root cause; Condor did not yet
support the delegated sessions at that time.

Tests were performed over the wide area network, compared to the loopback
device, and the following results were obtained (all times are in seconds):

\begin{center}
\begin{tabular}{ l | l | l }
\hline \hline
                    &   WAN  & loopback \\
\hline
ping time               &   0.15 & 0.00006 \\
bare GSI authentication (client)    &   0.43 & 0.15 \\
bare GSI authentication (server)    &   0.74 & 0.15 \\
CEDAR session setup with GSI (client)   &   1.0  & 0.15 \\
CEDAR session setup with GSI (server)   &   1.2  & 0.16 \\
   \hline
\end{tabular}
\end{center}

The network latencies are a performance killer for single threaded products
like the Condor daemons.

The first obvious bottleneck is the Condor daemon that keeps the state of the
resource pool (the collector), since it does a lot of communication.  In the US
CMS setup, each glidein initiates four sessions at startup.  US CMS mitigated
the collector scalability problem by deploying multiple daemons, which worked
great since collectors can be arranged in a tree fashion.  Using a tree of 1+70
collectors the system was able to handle first 10000 and then 25000 glideins
over a WAN, with a global glidein turnaround rate of up to 2.5Hz.

The other Condor daemon that is connection-intensive is the schedd, as it has
to handle the connections with the startds when the jobs start and end.  While
the schedds can also be replicated, they cannot be arranged in a tree
structure; each schedd has to be independent.  As such, from the usability
point of view, it is desirable to use a single schedd.

The schedd scalability problem was addressed by using the delegated sessions.
In this setup, only the collector needs to fully authenticate the parties (i.e
the schedd and startds) and establish a security session. The schedd is then
given the security session of the matched startd during the matchmaking
process, thus making communication between the schedd and the startd very
efficient.  With this setup, the schedd easily handled 25000 running jobs in
the above mentioned test.

\section{Conclusion}

This paper demonstrated how thoughtful network session management can
enhance the performance of a large-scale distributed system.  The
demands of establishing and resuming a secure channel across
administrative domains, multiple distributed services, and wide-area networks
created the challenges that motivated the development of CEDAR.  The
mechanisms in CEDAR's design, outlined in this paper, have played a key
role in allowing the Condor High Throughput Computing system to meet the
large and dynamic workload of the US CMS experiment.

\section*{References}
\bibliography{everything}

\end{document}